# FEAR: A Fuzzy-based Energy-Aware Routing Protocol for Wireless Sensor Networks

Ehsan Ahvar[1], Alireza Pourmoslemi[2], Md. Jalil Piran[3]

[1]Department of Information Technology & Communication, Payame Noor University, Iran
ehssana2000@yahoo.com

[2]Department of Mathematics, Payame Noor University, Iran
pourmoslemy@yahoo.com

[3]MIEEE, CSE Department, Jawaharlal Nehru Technological University, India
piran.mj@gmail.com

**Abstract**

Many energy-aware routing protocols have been proposed for wireless sensor networks. Most of them are only energy savers and do not take care about energy balancing. The energy saver protocols try to decrease the energy consumption of the network as a whole; however the energy manager protocols balance the energy consumption in the network to avoid network partitioning. This means that energy saver protocols are not necessarily energy balancing and vice versa. However, the lifetime of wireless sensor network is strictly depending on energy consumption; therefore, energy management is an essential task to be considered. This paper proposes an energy aware routing protocol, named FEAR, which considers energy balancing and energy saving. It finds a fair trade-off between energy balancing and energy saving by *fuzzy set* concept. FEAR routing protocol is simulated and evaluated by *Glomosim* simulator.

## 1. INTRODUCTION

A new class of networks has appeared in the last few years: the so-called Wireless Sensor Network (WSN). A WSN is a set of small autonomous systems, called sensor nodes (also known as motes), which communicate wireless and cooperate to solve at least one common application. These nodes have to collaborate to fulfill their tasks as usual a single node is incapable of doing so. The sensor networks are included of many sensor nodes, which are deployed with high density very close to the sensing environment or inside it, as the sensor environment is a field where the sensors scattered to sense it and collect the required data. The tiny sensor nodes, which consist of several major components as sensor, analog-to-digital convertor (ADC), processor, storage, transmitter, power unit and some other components based on their task such as location finding system, mobilizer and power generator, are able to interact with the phenomena and collect data, and forward the received data as well as a router.

Sensor nodes in the sensing field coordinate among themselves to produce high-quality information about the physical environment. These sensors have the ability to communicate either amongst each other or directly to an external base-station (BS). A base-station may be a fixed node or a mobile node capable of connecting the sensor network to an existing communications infrastructure or to the Internet where a user can access to the reported data. Many routing algorithms have been developed for sensor and ad hoc networks [1-18]. All of these routing protocols can be classified according to the network structure as flat, hierarchical, or location-based. In flat networks, all nodes play the same role while hierarchical protocols aim at clustering the nodes so that cluster heads can do some aggregation and reduction of data in order to save energy. Location-based protocols utilize the position information to relay the data to the desired regions rather than the whole network.

Energy consumption in sensor networks is a striking factor, it is because of that the batteries carried by each mobile node have limited power supply, processing power is limited, which in turn limits the quality and quantity of services and applications that can be supported by each node. In sensor networks, the nodes play multiple roles of data providing, data processing and data routing. The power required for data sensing varies to the kind of application.

Energy expenditure in data processing is much lower than data transmission, so power consumption is the most considerable challenge in sensor networks. Routing in WSNs is very challenging due to the inherent characteristics such as energy and bandwidth limitations that distinguish these networks from other wireless networks like mobile ad hoc networks or cellular networks. This paper addresses to design an energy-aware routing protocol for flat structure wireless sensor networks. The proposed protocol, FEAR, tries to save and balance energy consumption in network. It finds optimal route in energy level and hop-count both. Routing decisions in FEAR are based on the distance to the Base Station as well as remaining energy of nodes on the path towards the base station.

The rest of the paper is organized as follows, in section 2, we discuss our motivation, in section 3, we introduce the fuzzy set, we present our proposed methodology in section 4, and then we show the details and result of our simulation in section 5 and 6, finally in section 7 we conclude our paper.

## 2. MOTIVATION

Routing has a significant influence on the overall WSN lifetime, and providing an energy efficient routing protocol remains an open research issue [9]. Most of energy aware routing protocols are designed to save total energy consumption. They usually find the shortest path between Source and Sink to reduce energy consumption. In our opinion, an energy saver protocol that balances energy consumption is better than a poor energy server protocol. To find the best route with respect to the "shortest path" may lead to network partitioning. On the other hand, if the attention is only paid to the energy balancing, may cause the found route is a long path and the network lifetime decreased as well. On the other hand, finding best route only based on energy balancing consideration may lead to long path with high delay and decreases network lifetime. [22]

The SEER is a simple energy efficient routing protocol [11]. It tries to reduce number of transmissions. But it has poor idea about energy management and energy balancing. On the other hand, the LABER routing protocol [1] tries to balance energy consumption. But it has some impressive problems such as low accuracy in updating of energy and high control overhead. In the LABER, the Acknowledgement packet only is forwarded to previous sender, but the other neighbors cannot update energy level of the sender of Acknowledgement; low accuracy. Moreover, the Acknowledgment packet is an extra control overhead. This paper designs an energy-aware routing protocol for Wireless Sensor Networks to balance and save energy by fuzzy set technique. It also updates energy without Acknowledgment packets to increase accuracy and decrease control overhead.

## 3. INTRODUCTION TO FUZZY SET

The theory of fuzzy sets was introduced by Prof. L. Zadeh in 1965 [19]. After the pioneering work of Prof. Zadeh, there has been a great effort to obtain fuzzy analogues of classical theories. Fuzzy set theory is a powerful tool for modeling uncertainty and for processing vague or subjective information in mathematical models. Their main directions of development have been diverse and its applications to the very varied real problems. The notion central to fuzzy systems is that truth values (in fuzzy logic) or membership values (in fuzzy sets) are indicated by a value on the range [0, 1], with "0" and "1" representing absolute Falseness and absolute Truth respectively. Some concepts of fuzzy set theory are listed as follows [21]:

*Definition 3-1*: Let X be some set of objects, with elements noted as x.

*Definition 3-2*: A fuzzy set A in X is characterized by a membership function mA(x) which maps each point in X onto the real interval [0,1]. As mA(x) approaches 1, the "grade of membership" of x in A increases.

*Definition 3-3*: A is EMPTY iff for all x, $m_A(x) = 0$
*Definition 3-4*: A = B iff for all x: $m_A(x) = m_B(x)$
*Definition 3-5*: $m_{A'} = 1 - m_A$.
*Definition 3-6*: A is CONTAINED in B iff $m_A \leq m_B$.
*Definition 3-7*: C = A UNION B, where: $m_C(x) = MAX(m_A(x), m_B(x))$.
*Definition 3-8*: C = A INTERSECTION B where: $m_C(x) = MIN(m_A(x), m_B(x))$.
*Definition 3-9:* Given a fuzzy set $\tilde{A}$, the alpha-cut (or lambda cut) set of $\tilde{A}$ is defined by

$$A_\alpha = \{x | mA(x) \geq \alpha\}$$

## 4. FEAR ROUTING PROTOCOL

FEAR protocol is a reactive protocol which employs a lazy approach whereby nodes only discover routes to destination only when needed (on-demand). FEAR consumes much less bandwidth than proactive protocols such as Destination-Sequenced Distance-Vector (DSDV) protocol, but the delay in determining a route can be substantially large. FEAR protocol is an energy aware routing protocol, which consists of three major steps:

- Neighbor discovery
- Forwarding data
- Energy Update

Below we explain the steps in details;

### 4.1 Neighbor discovery

The Sink or the Base Station (BS) initializes the network by flooding the network with a broadcast message. Each node that receives the initiate packet, adds an entry to its "Neighbor Table" including "Neighbor ID", "Energy Level" and "Hop Count" to its "Neighbor Table". Then, the node make some changes on broadcast message such as (1) it increments the "Hop Count" field, (2) it changes the "Source Address" field to its address and (3) it changes the "Energy Level" field to its energy level and then retransmits it. Every node in the network retransmits the broadcast message only once, to all of its neighbors. When the initial broadcast message has been flooded through the network, each node knows hop count and energy level of its neighbors.

The Sink node periodically sends a broadcast message through the network; so that nodes add new neighbors joined the network to the "Neighbor Table" and remove neighbors that have failed to be an active member of the network.

### 4.2 Forwarding data

When a node observes an event, it initiates a routing process. One of the most challenges in the reactive protocols is how to select the next hop. However, in this paper we propose a new scenario to solve it. The proposed protocol, FEAR, is based on fuzzy set technique. It consists of two fuzzy sets; A and B.

$A$ is the fuzzy set of all neighbors' energy levels:

$$A = \{e_1, e_2, ..., e_n\}$$

$A$ has a membership function, mA($e_i$) which can be defined as below:

$$mA(e_i) = \lambda e_i, \quad 1 \leq i \leq n. \tag{1}$$

where $\lambda$ is a control parameter to limit energy factor in [0,1] interval and $e_i$ is energy level of (i)$^{th}$ neighbor.

Let $\alpha$ be obtained from the following formula:

$$\alpha = \frac{\sum_{i=1}^{n} mA(e_i)}{n} \quad \text{Then} \quad A_\alpha = \{e_i | mA(e_i) \geq \alpha\} \tag{2}$$

where $\alpha$ is energy threshold and $A_\alpha$ ($\alpha$-cut) is used to remove the neighbors with unacceptable energy level.

and $B$ is the fuzzy set of all neighbors' hop counts with membership function mB($h_i$).

$$B = \{h_1, h_2, ..., h_n\}$$

$$mB(h_i) = 1 - \frac{h_i}{MaxHop}, \quad 1 \leq i \leq n \tag{3}$$

Where $n$ is the number of neighbors, and $h_i$ is hop count of (i)$^{th}$ neighbor.

Now, we define following decision maker equation:

$$C(i) = \begin{cases} mA_\alpha(e_i) \times mB(h_i) & \lambda e_i > \alpha \\ 0 & \lambda e_i \leq \alpha \end{cases}, \text{ Where } 1 \leq i \leq n. \tag{4}$$

However, the neighbor with maximum amount of C is selected as the next hop.

As used in equation (3), "MaxHop" is the estimation of the longest possible route in the network which plays an essential role on the FEAR's decisions. This paper tries to select optimal "MaxHop". Very large amount of MaxHop increases effect of "hopcount" factor compared with energy on equation (1) and vice versa.

To compute MaxHop some methods are proposed and discussed in this paper. These methods are classified into two different categories: Dynamic and static methods. For all proposed methods the sensor network with size of X*Y and nodes with Radio range of R is supposed.

First method computes maximum hop count based on the longest available distance and radio range of R, Fig.1 (a). It finds maximum possible distance between two nodes and then computes the "MaxHop" based on the following equation:

$$MaxHop = \sqrt{(X^2) + (Y^2)} / R \qquad (5)$$

However, the previous method considers the longest available path between Source and Destination and computes maximum hop count based on longest available distance. But the distance between two nodes or range of one hop was assumed with fix rang of R. This assumption may not be feasible. There is no accrued knowledge about the distance between two nodes.

To obtain an accurate "MaxHop", R is changed to R/2, Fig.1 (d), therefore:

$$MaxHop = \sqrt{(X^2) + (Y^2)} / (R/2) \qquad (6)$$

Also consider a network consisting of N nodes. It is possible to have a path between source and destination with N-2 hops Fig.1 (b). In this condition the "MaxHop = N-2".

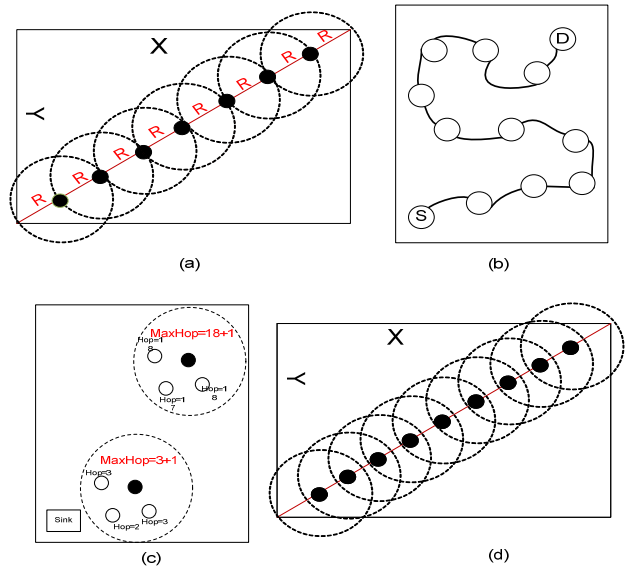

Figure 1.  Different shapes of finding MaxHop.

Generally, static "MaxHop" method is not too accurate. "MaxHop" is difficult to find and also constant amount of "MaxHop" leads to increase effect of "hopcount" factor for the nodes close to the sink and decrease effect of this factor for the further nodes.

Assume all nodes have equal energy levels. Nodes which are cloase to the Sink, HopCount<<MaxHop and Hop factor is almost one. But for the nodes located in a long distance to the sink, HopCount~MaxHop and Hop Factor will be about zero. Therefore, at different locations of network the effect of Hop factor is variable and constant "MaxHop" can not support a fair tradeoff between Hop and Energy Factors, Fig.2.

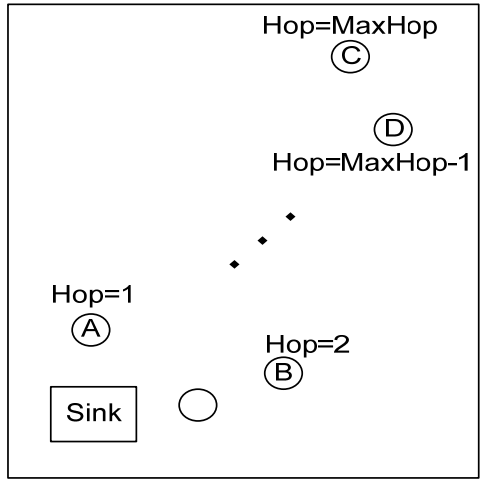

Figure 2.  Static method scheme

Because of weaknesses of static methods, the Dynamic Method is proposed Fig.1 (c). In this method "MaxHop" is computed prior to each forwarding step. In this scenario, each node looks at its 'Neighbor Table" and finds out the biggest hop count. Then; MaxHop = Biggest hop count + 1.

### 4.3 Energy Update

All the neighbors of the sender node receive the forwarded data packet via the overhearing technique. Then, they update the energy level of sender node in their "Neighbor Table" by the piggybacking technique. Nodes might be used by more than one neighbor for routing, in this case the energy value stored in the "Neighbor Tables" of both the node's neighbors will be completely accurate by the overhearing technique.

The operation of FEAR protocol can be summarized as follows:

- The sink initializes the network by flooding the network with a broadcast message.
- Nodes add all their neighbors' information to their neighbor tables.
- The node with highest $C(i)$ is selected and data packet is forwarded to it.
- The energy level of data sender node is updated by its neighbors.

FEAR routing protocol does not need any energy message because the energy level of each sender is updated into its "Neighbor Table" by overhearing and piggybacking techniques automatically. Also, the sink node sends a broadcast message through the network, so that nodes can add new neighbors that joined the network to the "Neighbor Table" and remove neighbors that have failed. However, the sending rate of broadcast message through the network is related to the nodes' mobility. In contrary to the networks with high mobility nodes, the networks with non mobility nodes do not need to send broadcast message.

### 5. SIMULATION METHODOLOGY

For simulation work two features of the proposed protocol should be considered: (1) FEAR with dynamic "MaxHop", known as Dynamic FEAR (D-FEAR) and (2) FEAR with Constant "MaxHop", known as Static FEAR (S-FEAR). The constant "MaxHop" is obtained based on radio range of R/2, and if maximum hop count is bigger than "MaxHop", then maximum hop count will be assumed "MaxHop". To compare the routing protocols, a parallel discrete event-driven simulator, *GloMoSim*, is used. GloMoSim is a simulation tool for large wireless and wired networks [20].
Table 1 describes the detailed setup for our simulator.

| SIMULATION-TIME | 1200 SECOND |
|---|---|
| TERRAIN-DIMENSIONS | 1000m*1000m |
| NUMBER-OF-NODES | 200, 300, 500, 1000, 2000 |
| NODE-PLACEMENT | Uniform/Random |

| MOBILITY | NONE |
|---|---|
| NUMBER OF EVENTS (Sources) | 100 |
| TEMPARATURE | 290.0 (in K) |
| RADIO-BANDWIDTH | 2000000(in bps) |
| RADIO-TX-POWER | 5.0 (in dBm) |
| ENERGY- TRANSMIT-LEVEL | 0.0002 (in mW) |
| MAC-PROTOCOL | 802.11 |
| NETWORK-PROTOCOL | IP |
| PROPAGATION-PATHLOSS | FREE-SPACE |
| RADIO-TYPE | RADIO-ACCNOISE |

**Table1. Simulation setting**

## 6. SIMULATION RESULT

In this section we evaluate and compare the various routing schemes. The interested performance measures in this study are: (a) Average energy consumption of transmission (in mW); (b) Energy balancing. The variables are: number of nodes and node placement.

The simulation of the protocol started with a broadcast message. We select 100 Sources to send data packets to the Base Station during the simulation. The Sources are selected randomly in different times. Each Source generates a 512-bit data packet and forwards it through the network. Simulations are performed to evaluate the network lifetime achieved by each protocol. At the beginning of simulation, the transmission energy level of each node was 0.0002 mW. We design four different tests to evaluate protocols as follows:

**Test 1**: *The time until the first neighbor of station fails.* The first failure of station neighbor is related to energy management. Energy balancer protocols should have better failure time than the other protocols. Therefore, this test evaluates energy management of each protocol.

**Test 2**: *The Number of fails.* This test computes number of fails because of their depleting their energy sources. The protocol with the lowest number of fails is the best in energy factor.

**Test 3:** *Percentage of active neighbors of the station at the end of simulation.* This test shows protocols' ability to keep the station connected.

**Test 4:** *The average remaining energy of all the nodes in the network, at transmission mode.* This test is designed to find which protocol is more energy saver than the other.

After the simulation the following results are achieved:

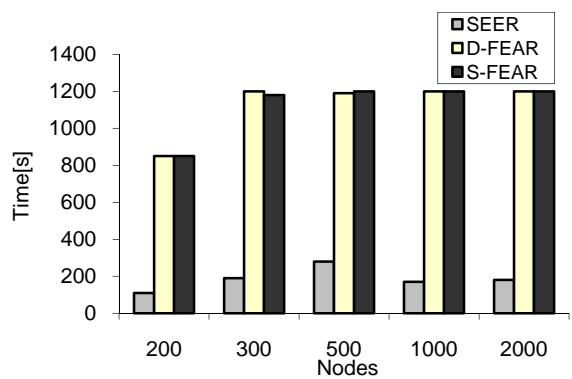

Figure 3. Time at which the first neighbor of Station fails due to depleting its energy source. The nodes are randomly distributed over the sensor area.

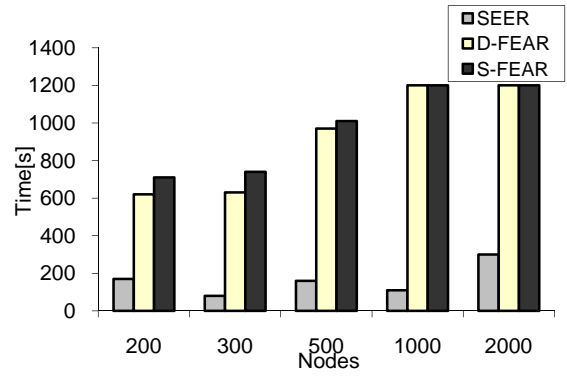

Figure 4. Time at which the first neighbor of Station fails due to depleting its energy source. The nodes are uniformly distributed over the sensor area.

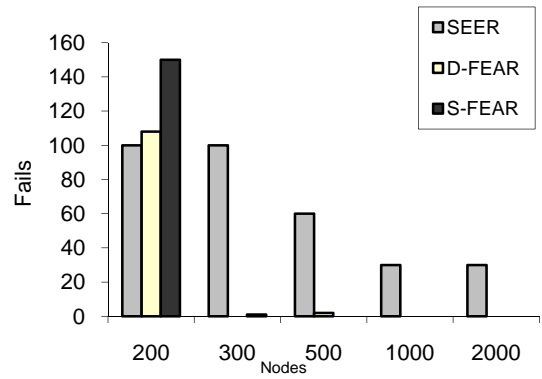

Figure 5. Number of fails at the end of simulation. The nodes are randomly distributed over the sensor area.

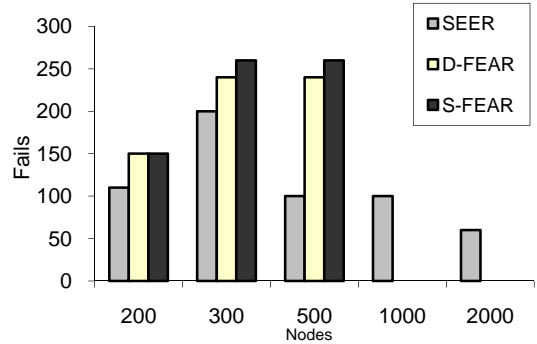

Figure 6. Number of fails at the end of simulation. The nodes are uniformly distributed over the sensor area.

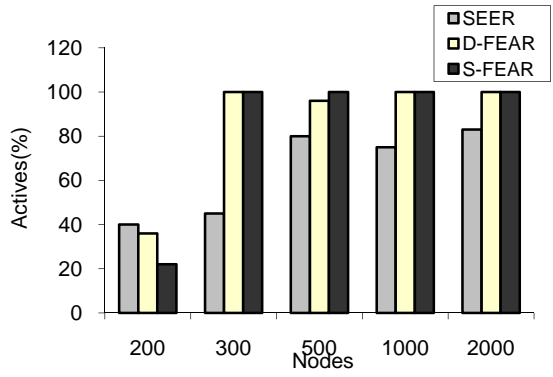

Figure 7.  Percentage of active neighbors of Station at the end of simulation. The nodes are randomly distributed over the sensor area.

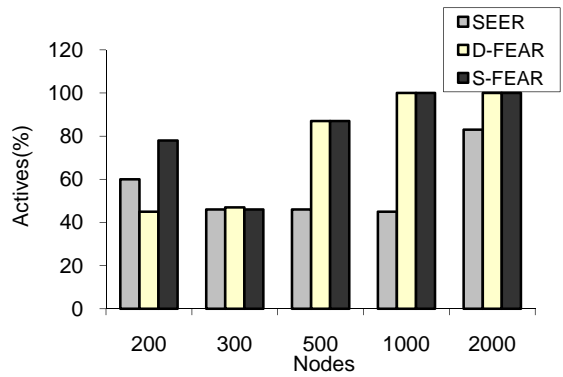

Figure 8.  Percentage of active neighbors of Station at the end of simulation. The nodes are uniformly distributed over the sensor area.

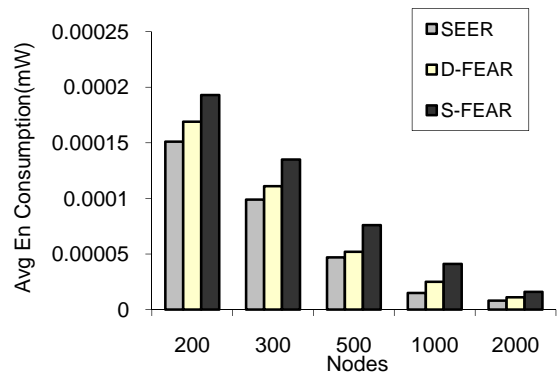

Figure 9.  Average energy consumption(mW) in transmission mode. The nodes are randomly distributed over the sensor area.

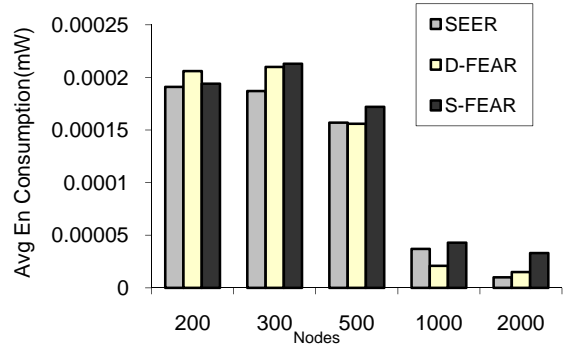

Figure 10. Average energy consumption(mW) in transmission mode. The nodes are uniformly distributed over the sensor area.

Results of Test 1 (figure3 and figure 4) show that there is a considerable discrepancies between the SEER and FEAR protocols in the time at which the first neighbor of the Base Station fails due to its depleting energy source. D-FEAR and S-FEAR are more optimal than SEER protocol in energy management. This is due to the fact that FEAR sends data packet along an optimum energy-balanced path. But there is no considerable difference between D-FEAR and S-FEAR Results of Test 2 (figures 5 & 6) show that at the end of simulation, FEAR has very low fails compared with the SEER. Consequently, our proposed protocol has a acceptable performance in high-density networks.

Test 3 (figures 7 & 8) shows the percentage of active neighbors of the Base Station at the end of simulation. One can see that D-FEAR and S-FEAR have better performance than SEER especially in high density networks.

Test 4 (figures 9 & 10) shows that there is no visible difference in energy consumption between SEER, D-FEAR and S-FEAR.

As yielded desired and acceptable results by the abovementioned four tests, it is salient that FEAR has fair performance in "Energy Balance" aspect. So, it's more suitable for wireless sensor networks. According to the test results it is obvious that the proposed FEAR routing protocol increase the network lifetime in compare to the SEER. Also, the performance of the S-FEAR has been improved by presenting D-FEAR.

## 7. CONCLUSION

In this paper we proposed a new energy saver/balancer routing protocol, named FEAR. We used fuzzy technique to balance energy. The proposed routing protocol was simulated and compared to the traditional SEER. Results showed that FEAR routing protocol is better than SEER in energy balancing. FEAR protocol has very low fails. It is more balancer than the SEER routing protocol especially in high-density networks.